\documentclass[aps,pra,twocolumn,10pt,superscriptaddress,nofootinbib,balancelastpage]{revtex4-2} 
\usepackage[latin1]{inputenc}
\usepackage{amsmath,amssymb}
\usepackage{mathrsfs} 
\usepackage[capitalise]{cleveref}
\usepackage{braket}
\usepackage{graphicx,color,colortbl}
\usepackage{booktabs}
\usepackage{seqsplit}
\usepackage{lipsum}
\usepackage{dsfont}

\allowdisplaybreaks

\newcommand{\R}{\mathbb{R}}%
\newcommand{\dif}{\mathrm{d}}%
\newcommand{\Kronecker}[2]{\delta_{#1#2}}%
\newcommand{\Nabla}{\vec{\nabla}}%
\newcommand{\Laplace}{\boldsymbol{\triangle}}%
\newcommand{\norm}[1]{\lVert#1\rVert}%

\newlength{\myl}%
\newcommand{\INT}[3]{\settowidth{\myl}{$\displaystyle\int_{#1}^{#2}$}{\int_{#1}^{#2}\;\;\;\hspace{-\the\myl}\dif #3}\,}% Integrale in abgesetzten Gleichungen
\newcommand{\TINT}[3]{\settowidth{\myl}{$\displaystyle\int_{#1}^{#2}$}{\int_{#1}^{#2}\;\;\;\;\;\,\hspace{-\the\myl}\dif #3}\,}% Integrale in Flie{\ss}textgleichungen
\newcommand{\EINT}[3]{\settowidth{\myl}{$\int_{#1}^{#2}$}{\int_{#1}^{#2}\;\;\;\,\hspace{-\the\myl}\dif #3}\,}% Integrale in Exponenten

\makeatletter
\newcommand\footnoteref[1]{\protected@xdef\@thefnmark{\ref{#1}}\@footnotemark}
\makeatother

\begin{document}

\title{Active Brownian particles in external force fields: field-theoretical models, generalized barometric law, and programmable density patterns}

\author{Jens Bickmann}
\thanks{These two authors contributed equally.}

\author{Stephan Br\"oker}
\thanks{These two authors contributed equally.}

\author{Raphael Wittkowski}
\email[Corresponding author: ]{raphael.wittkowski@uni-muenster.de}
\affiliation{Institut f\"ur Theoretische Physik, Center for Soft Nanoscience, Westf\"alische Wilhelms-Universit\"at M\"unster, 48149 M\"unster, Germany}

\date{\today}

\begin{abstract}
We investigate the influence of external forces on the collective dynamics of interacting active Brownian particles in two as well as three spatial dimensions. Via explicit coarse graining, we derive predictive models that are applicable for space- and time-dependent external force fields. We study these models for the cases of gravity and harmonic traps. In particular, we derive a generalized barometric formula for interacting active Brownian particles under gravity that is valid for low to high concentrations and activities of the particles. Furthermore, we show that one can use an external harmonic trap to induce motility-induced phase separation in systems that, without external fields, remain in a homogeneous state. This finding makes it possible to realize programmable density patterns in systems of active Brownian particles. Our analytic predictions are found to be in very good agreement with Brownian dynamics simulations.
\end{abstract}
\maketitle

\section{Introduction}
Active Brownian particles (ABPs) are particles that undergo Brownian motion together with constant self-propulsion \cite{Elgeti2015,Speck2016,Zttl2016,BechingerEA16}. Under the influence of external fields, they can show a variety of effects \cite{BechingerEA16,katuri2016artificial,Stark2016}. This includes anomalous sedimentation profiles under gravity \cite{Wang2014,ginot2015nonequilibrium,vachier2019dynamics}, self-induced polar ordering \cite{campbell2013gravitaxis,wolff2013,locatelli2015active,campbell2017helical,Ginot2018,tenHagen2014gravitaxis}, trapping \cite{TailleurCates2009,pototsky2012,tenHagen2014gravitaxis}, superfluidity \cite{SuperFluidBacteriaLopez15,Takatori17}, effective diffusion coefficients \cite{Palacci2010, Maggi2013, wolff2013, koumakis2014, Wang2014, Hermann2018}, and self-organized fluid pumps \cite{Hennes2014}. In addition, ABPs show an accumulation at repulsive walls or interfaces \cite{vladescu2014filling,yang2014}. Such walls or interfaces can be described via an external force field \cite{mones2015}. A force field can also be used to determine properties like pressure in far-from-equilibrium systems \cite{solon2015pressure, Ginot2018, Solon2018}. Furthermore, the nonequilibrium dynamics of active particles in external fields is very important for future applications of such particles in medicine and materials science, where active particles can, e.g., perform drug delivery \cite{koumakis2013, Ma2015, Li2017, Santiago2018,nitschkeSW2021collective} and form active materials with exceptional properties \cite{ActiveMaterials2021book}, respectively. It has been demonstrated that the control over the particles that is needed for such applications can be well achieved via external fields \cite{yang2018colloidal,nitschkeSW2021collective}. The active agents can be either artificial self-propelled microparticles \cite{Rao2015,Wu2016,Xu2016,Guix2018,Chang2019,PachecoJerez2019} or motile microorganisms \cite{SchwarzLinek2016,Chen2017,Andac2019}. Both are frequently and successfully described as ABPs, including even run-and-tumble particles like \textit{Escherichia coli} bacteria \cite{Tailleur2008,Cates2013,Liu2017,Andac2019}. 
 
Despite the importance of the behavior of ABPs under an external force, there exist only a few, and often very specific, theoretical models for the nonequilibrium dynamics of ABPs in external force fields. Sedimentation profiles for noninteracting ABPs in two and three spatial dimensions are derived in Ref.\ \cite{Ginot2018} and further compared with experimental data. The authors of this work found a qualitative agreement with the traditional barometric formula only after a certain height. In Ref.\ \cite{vachier2019dynamics}, a coarse graining in Fourier space with long-wavelength approximations gives a model for noninteracting ABPs under gravity in three spatial dimensions and in the presence of a confining wall, resulting in a refined barometric formula in the steady-state limit. By neglecting translational diffusion, the steady-state equation for systems of noninteracting ABPs in two spatial dimensions under the effect of various external forces was solved analytically in Ref.\ \cite{wagner2017}. Reference \cite{Hermann2018} is even able to present an exact result for the steady state in systems of noninteracting ABPs in two dimensions under gravity and with translational diffusion, but considers only situations with fixed polarization of the particles' orientations. Another work \cite{tenHagen2014gravitaxis} investigates `L'-shaped particles and their polar ordering in gravitational fields based on theory and experiments, but their method does not include a field-theoretical description. ABPs confined by two-dimensional optical traps are investigated in Ref.\ \cite{pototsky2012}, where a corresponding dynamical density functional theory (DDFT) \cite{teVrugtLW2020DDFTreview} for interacting ABPs, based on a mapping to passive colloidal particles, was derived. Further work on DDFT for microswimmers in confinement can be found in Refs.\ \cite{Menzel2016j,Hoell2017,hoell2019multi}. The DDFT approach, however, has the consequence that the model is only valid at low densities and low activities of the particles. In Ref.\ \cite{MariniBettoloMarconi2015}, a statistical field theory for a specific model system with arbitrary external forces was derived using the unified colored-noise approximation \cite{JungH1987UCNA}.

Altogether, these field theories have in common that they are either nonlocal or nonpredictive and include strong approximations. The locality of a field theory allows for a more accessible interpretation of terms and a simpler numerical implementation. A predictive theory enables the connection of microscopic parameters of the considered system to its macroscopic description, hence giving deeper insights into the system's dynamics. Also, in many field theories (e.g., DDFT approaches) the theory is very limited in its range of application, and often valid only at low particle densities and therefore not well suited to describe dense and interacting collectives of ABPs. A local predictive field theory that considers arbitrary external force fields, while still accounting for the full interaction between the particles, is still missing and would be an important asset for the area of research.

In this article, we derive such a local predictive field theory for interacting ABPs in two and three spatial dimensions under the influence of external force fields that can be space- and time-dependent. As results, we present models that describe the dynamics of the particles up to the second and fourth orders in derivatives. We show that in the case of the $2$nd-order-derivatives model, external forces mimic advection. Using our models, we study the stationary states of the particles' dynamics for the case of two common types of external fields: gravity and harmonic traps. For the case of gravity, we observe deviations from the Boltzmann distribution due to an interplay of the particles' activity and interactions. For harmonic traps, we show that the ABPs can be induced to undergo motility-induced phase separation (MIPS) \cite{MIPS} at desired locations. We verify this result by comparing with Brownian dynamics simulations. Our analytical predictions and the simulation data are found to be in very good agreement.

This article is structured as follows: In section \ref{sec:Methods}, we derive the models and describe the simulation setup. We apply our model to the cases of gravity fields and harmonic traps and discuss the results in section \ref{sec:results}. Finally, we conclude in section \ref{sec:Conclusion}.

\section{\label{sec:Methods}Methods}
\subsection{\label{sec:derivation}Theoretical model}
To derive a local, predictive field theory for interacting ABPs in external force fields, we consider systems of $N$ similar, spherical ABPs in two (2D) and three (3D) spatial dimensions. The $i$-th ABP is described using its center-of-mass position $\vec{r}_i(t)$ and normalized orientation vector $\hat{u}_i(t)$, which are functions of time $t$. Its motion is influenced by an external force field $\vec{F}_{\mathrm{ext}}(\vec{r}_i, t)$ that can depend on the particle position and time. The translational motion of the ABPs considered here is given by the Langevin equation 
\begin{align}
\dot{\vec{r}}_i(t) &=\vec{\xi}_{\mathrm{T},i} + v_0\hat{u}_i +  \beta D_\mathrm{T}\vec{F}_{\mathrm{int}, i}(\{\vec{r}_i\}) + \vec{v}_\mathrm{ext}(\vec{r}_i, t),
\label{eqn:r-LangevinAllgemein}%
\end{align}
which holds for 2D and 3D. 
A partial derivative with respect to time is denoted by an overdot. 
For the rotational motion of the ABPs, we use the Langevin equation 
\begin{align}
\dot{\phi}_i &= \xi_{\mathrm{R}, i}\qquad\qquad &\text{for 2D},
\label{eqn:phi-Langevin}\\
\dot{\hat{u}}_i &= \hat{u}_i \times \vec{\xi}_{\mathrm{R}, i}\qquad&\text{for 3D},
\label{eqn:u-Langevin}%
\end{align}
with the parametrization 
\begin{align}
\hat{u}_i(\phi) &= (\cos(\phi), \sin(\phi))^\mathrm{T} &\text{for 2D},\\
\hat{u}_i(\theta,\phi) &= (\cos(\phi)\sin(\theta), \sin(\phi)\sin(\theta), \cos(\theta))^\mathrm{T} &\text{for 3D}
\end{align}
of the orientation vector by polar (2D) and spherical (3D) coordinates, respectively. 
In Eqs.\ \eqref{eqn:r-LangevinAllgemein}-\eqref{eqn:u-Langevin}, the translational and rotational Brownian motion of the $i$-th particle is described by statistically independent Gaussian white noises $\vec{\xi}_{\mathrm{T}, i}(t)$ for translation (2D and 3D) and $\xi_{\mathrm{R}, i}(t)$ (2D) or $\vec{\xi}_{\mathrm{R}, i}(t)$ (3D) for rotation, respectively. These noises have zero mean and their correlations are given by $\braket{\vec{\xi}_{\mathrm{T}, i}(t_1)\otimes\vec{\xi}_{\mathrm{T}, j}(t_2)} = 2D_\mathrm{T}\Kronecker{i}{j}\mathds{1}_n\delta(t_1-t_2)$ for translation (2D and 3D) and $\braket{\xi_{\mathrm{R}, i}(t_1)\xi_{\mathrm{R}, j}(t_2)} = 2D_\mathrm{R}\Kronecker{i}{j}\delta(t_1-t_2)$ (2D) or $\braket{\vec{\xi}_{\mathrm{R}, i}(t_1)\otimes\vec{\xi}_{\mathrm{R}, j}(t_2)} = 2D_\mathrm{R}\Kronecker{i}{j}\mathds{1}_3\delta(t_1-t_2)$ (3D) for rotation with the ensemble average $\braket{\,\cdot\,}$, dyadic product $\otimes$, translational and rotational diffusion coefficients $D_{\mathrm{T}}$ and $D_{\mathrm{R}}$, respectively, Kronecker delta $\delta_{ij}$, and the $n\times n$-dimensional identity matrix $\mathds{1}_n$, where $n=2$ for 2D and $n=3$ for 3D. For the spherical particles, the Stokes-Einstein-Debye relation $D_\mathrm{R} = 3 D_\mathrm{T}/\sigma^2$ holds, where $\sigma$ is their diameter. Furthermore, $v_0$ denotes the propulsion speed of an individual ABP that is not affected by interactions or an external force, $\beta = 1/(k_\mathrm{B}T)$ is the thermodynamic beta with Boltzmann constant $k_\mathrm{B}$ and absolute temperature $T$, and $\vec{F}_{\mathrm{int}, i}(\{\vec{r}_i\})=-\sum_{j=1, j\neq i}^{N}\Nabla_{\vec{r}_i} U_2(\norm{\vec{r}_i-\vec{r}_j})$ is the particle-particle interaction force, where $\Nabla_{\vec{r}_i}$ denotes the del operator with respect to $\vec{r}_i$ and $U_2$ the pair-interaction potential of the particles. The propulsion speed that would solely originate from the external force can be written as $\vec{v}_\mathrm{ext}(\vec{r}_i, t) = \beta D_\mathrm{T}\vec{F}_{\mathrm{ext}}(\vec{r}_i, t)$ and constitutes the central object of our investigation. 

Equations \eqref{eqn:r-LangevinAllgemein}-\eqref{eqn:u-Langevin} correspond to the statistically equivalent Smoluchowski equation 
\begin{equation}
\begin{split}
\dot{\mathfrak{P}} & = \sum_{i=1}^N \Big( -v_0\hat{u}_i\cdot\Nabla_{\vec{r}_i}\mathfrak{P} + (D_\mathrm{T}\Laplace_{\vec{r}_i} + D_\mathrm{R} \mathfrak{R}_i^2 )\mathfrak{P}\\
&\qquad\quad\;\; - \Nabla_{\vec{r}_i}\cdot \big( (\beta D_\mathrm{T}\vec{F}_{\mathrm{int}, i}(\{\vec{r}_i\}) + \vec{v}_\mathrm{ext}(\vec{r}_i, t) ) \mathfrak{P}\big) \Big),
\end{split}\raisetag{3em}
\end{equation}
which describes the time evolution of the many-particle probability density $\mathfrak{P}(\lbrace\vec{r}_i\rbrace, \lbrace\hat{u}_i\rbrace, t)$ that can depend on the whole set of position vectors $\lbrace\vec{r}_i\rbrace$, the set of orientations $\lbrace\hat{u}_i\rbrace$, and time $t$. Here, the Laplace operator acting on the $i$-th particle is denoted as $\Laplace_{\vec{r}_i}\equiv\Nabla^2_{\vec{r}_i}$ and the rotational operator is given by $\mathfrak{R}_i = \frac{\partial}{\partial \phi_i}$ (2D) or $\mathfrak{R}_i = \hat{u}_i\times \frac{\partial}{\partial \hat{u}_i}$ (3D).
 
The further derivation of a field-theoretical model is based on the \textit{interaction-expansion method} \cite{RW, BickmannW2020, BickmannW2020b,BickmannBJW2020Chiraliry}. By integrating over all degrees of freedom except for those of one particle, renaming them as $\vec{r}$ and $\hat{u}$, and multiplying with $N$, an equation for the one-particle density field 
\begin{align}
\varrho(\vec{r}, \phi, t) &= N\Bigg(\prod_{\begin{subarray}{c}j=1\\j\neq i\end{subarray}}^{N}\int_{\R^2}\!\!\!\!\dif^2r_j\int_{0}^{2\pi}\!\!\!\!\!\!\dif \phi_j\Bigg)\, \mathfrak{P}\bigg\rvert_{\begin{subarray}{l}\vec{r}_i=\vec{r},\\\phi_i=\phi\end{subarray}}\!\!&\text{for 2D,}\label{eqn:2DVarRho-Projektor} \\
\varrho(\vec{r}, \hat{u}, t) &= N\Bigg(\prod_{\begin{subarray}{c}j=1\\j\neq i\end{subarray}}^{N}\int_{\R^3}\!\!\!\!\dif^3r_j\int_{\mathbb{S}}\!\dif^2u_j \Bigg)\, \mathfrak{P}\bigg\rvert_{\begin{subarray}{l}\vec{r}_i=\vec{r},\\\hat{u}_i=\hat{u}\end{subarray}} \!\!&\text{for 3D}\label{eqn:3DVarRho-Projektor}
\end{align}
can be obtained. We denote the surface of the unit sphere in 3D by $\mathbb{S}$.
 
Additional steps of the derivation include a Fourier expansion (2D) \cite{BickmannW2020} or spherical harmonics expansion (3D) \cite{BickmannW2020b} of the particles' pair-distribution function $g$, a gradient expansion \cite{Yang1976,Evans1979,EmmerichLWGTTG12} of integrals that contain the interaction force, an orientational expansion of $\varrho$ into Cartesian order-parameter tensors \cite{teVrugt19}, and a quasi-stationary approximation (QSA) \cite{RW,BickmannW2020,BickmannW2020b} of the resulting coupled equations for the Cartesian order-parameter fields.

The pair-distribution function $g$ relates the two-particle density $\varrho^{(2)}(\vec{r}, \vec{r}', \hat{u}, \hat{u}', t)$ to one-particle densities $\varrho(\vec{r}, \hat{u}, t)$:
\begin{equation}
\varrho^{(2)}(\vec{r}, \vec{r}', \hat{u}, \hat{u}', t) = g(\vec{r}, \vec{r}', \hat{u}, \hat{u}', t)\varrho(\vec{r}, \hat{u}, t)\varrho(\vec{r}', \hat{u}', t).
\end{equation}
Both correlation functions depend on the number of spatial dimensions and are in general unknown. However, for stationary states, where $g$ has translational and rotational invariance and is time-independent, analytic representations of $g$ exist in 2D \cite{Jeggle2019} and 3D \cite{Broeker2019}. To apply these representations, we assume that $g$ is approximately translationally and rotationally invariant and time-independent on the scale of the particles' interaction length and angular relaxation time. For short-range interactions and weak spatial and temporal changes of the external force, these assumptions are well justified.

After performing the aforementioned derivation steps, the dynamics of the system is described in terms of the local particle number density 
\begin{align}
\rho(\vec{r}, t) &= \int_0^{2\pi}\!\!\!\!\!\!\dif\phi\, \varrho(\vec{r}, \hat{u}(\phi), t)\qquad &\text{for 2D},\\
\rho(\vec{r}, t) &= \int_{\mathbb{S}}\!\dif^2u\, \varrho(\vec{r}, \hat{u}, t)\qquad\quad\, &\text{for 3D}.
\end{align}
We restrict the gradient expansion to terms of maximal order two in derivatives and, since we want to consider weak external fields, we neglect terms of second or higher order in $\vec{v}_\mathrm{ext}$. The resulting model is an advection-diffusion equation
\begin{equation}
\dot\rho = \Nabla\cdot( -\vec{v}_{\mathrm{ext}}\rho + D(\rho)\Nabla\rho) \label{eqn:2ndOrderAdvectionDiffusion}
\end{equation}
with density-dependent diffusion coefficient 
\begin{equation}
D(\rho) = D_\mathrm{T} + a_0 + a_1 \rho + a_2 \rho^2.\label{eqn:DDdiffusion}
\end{equation}
In Eq.\ \eqref{eqn:2ndOrderAdvectionDiffusion}, the external force field mimics an advection velocity, and the coefficients $\lbrace a_i\rbrace$ in Eq.\ \eqref{eqn:DDdiffusion} depend on the number of spatial dimensions of the system. These coefficients are related to microscopic parameters of the system by equations that are given in Appendix \ref{app:A}. In the limit of dilute suspensions, the density-dependence of $D(\rho)$ can be neglected and Eq.\ \eqref{eqn:2ndOrderAdvectionDiffusion} obtains the same form as for passive particles, but with a different diffusion coefficient $D_\mathrm{T} + a_0$. This provides a mapping of the ABPs to passive particles with an effective diffusion coefficient \cite{Palacci2010}.
When ignoring the advection term $-\Nabla\cdot(\vec{v}_{\mathrm{ext}}\rho)$, Eq.\ \eqref{eqn:2ndOrderAdvectionDiffusion} becomes equivalent to Eqs.\ (20)-(22) in Ref.\ \cite{BickmannW2020} for 2D or Eqs.\ (21)-(23) in Ref.\ \cite{BickmannW2020b} for 3D. 
Equation \eqref{eqn:2ndOrderAdvectionDiffusion} constitutes the simplest model describing systems of ABPs under the influence of an external force. 

Although this $2$nd-order-derivatives model can predict the onset of MIPS, one might be interested also in a description of the further time-evolution of MIPS. For such a description, one needs at least four orders in derivatives, as they are present in models like Active Model B \cite{Wittkowski2014}, Active Model B + \cite{Tjhung2018}, and predictive field theories proposed in Refs.\ \cite{BickmannW2020} (2D) and \cite{BickmannW2020b} (3D). 
By truncating the gradient expansion at $4$th-order derivatives and performing a QSA, we obtain a field theory that extends the phase-field models of Refs.\ \cite{BickmannW2020,BickmannW2020b} towards an external force field. The dynamic equation for the density field in this field theory is
\begin{equation}
\begin{split}
\dot{\rho} &= -\Nabla\cdot(\vec{J}^{(\mathrm{int})} + \vec{J}^{(\mathrm{ext})}).
\end{split}
\label{eq:Modell4O}%
\end{equation}
Here, $\vec{J}^{(\mathrm{int})}$ denotes the density current of ABPs under no external force, which is given by Eq.\ (25) in Ref.\ \cite{BickmannW2020} for 2D and Eq.\ (26) in Ref.\ \cite{BickmannW2020b} for 3D and can also be found in Appendix \ref{app:Jint} of the present article. The current $\vec{J}^{(\mathrm{ext})}$ arises from the external force field. 
When we again neglect terms of second or higher order in $\vec{v}_{\mathrm{ext}}$, the current $\vec{J}^{(\mathrm{ext})}$ reads
\begin{align}
\begin{split}
\vec{J}^{(\mathrm{ext})} &= \vec{v}_{\mathrm{ext}}\rho + (b_1 + b_2\rho + b_3\rho^2)\Nabla\cdot(\vec{v}_{\mathrm{ext}}\otimes\Nabla\rho)\\
&\quad +(b_4+b_3\rho)(\vec{v}_{\mathrm{ext}}\cdot\Nabla\rho) \Nabla\rho,\label{eqn:4thOrderEXTcurrent}
\end{split}
\end{align}
where we use the notation $\vec{a}\cdot (\vec{b}\otimes\vec{c}) = (\vec{a}\cdot\vec{b})\vec{c}$. 
Microscopic expressions for the coefficients $\lbrace b_i\rbrace$ are given in Appendix \ref{app:A}. 

Note that the advection-diffusion model \eqref{eqn:2ndOrderAdvectionDiffusion} and the extended phase-field model \eqref{eq:Modell4O} of APBs in external fields can be applied for low to high particle densities and small to large activities. These models constitute the first main result of this article. 
While in the $2$nd-order-derivatives model the external force field occurs via an advection velocity $\vec{v}_\mathrm{ext}$, the $4$th-order-derivatives model has additional contributions of $\vec{v}_\mathrm{ext}$.

If one considers stationary states, from Eq.\ \eqref{eqn:2ndOrderAdvectionDiffusion} follows that the density field must obey the equation
\begin{equation}
D(\rho)\Nabla\rho = \vec{v}_\mathrm{ext}\rho. 
\label{eqn:StationaryEq}%
\end{equation}
Note that $\vec{v}_\mathrm{ext}$ can still be time-dependent, albeit only on the time scale of the relaxation time of the stationary state or slower. 
Equation \eqref{eqn:StationaryEq} allows to obtain stationary-state solutions for the density of interacting ABPs in a simple way. It is an extension of the equation $D(\rho)=0$ for the stationary state of interacting ABPs in the absence of external forces \cite{Bialk2013}.

\subsection{\label{ssec:SimDetails}Simulations}
In our simulations, we focus on the 2D case to keep the computational effort moderate. 
Numerical solutions of the stationary-state equation \eqref{eqn:StationaryEq} are obtained using the adaptive Runge-Kutta-Fehlberg 4(5) method with an accuracy goal of $10^{-10}$ \cite{Mathematica}.

When studying harmonic traps, we perform also Brownian dynamics simulations based on the Langevin equations \eqref{eqn:r-LangevinAllgemein} and \eqref{eqn:phi-Langevin} using a modified version of the software package LAMMPS \cite{Plimpton1995}. As interaction potential, we choose the purely repulsive Weeks-Chandler-Andersen potential \cite{WeeksCA1971} 
\begin{equation}
U_{2}(r)=
\begin{cases}
4\varepsilon \Big( \big( \frac{\sigma}{r} \big)^{12} - \left( \frac{\sigma}{r} \right)^{6} \Big) + \varepsilon, & \mbox{if } r \leq 2^{1/6} \sigma, \\
0, & \mbox{else}
\end{cases}
\label{eq:wca}%
\end{equation}
with the interaction energy $\varepsilon$ and particle diameter $\sigma$. 
To incorporate a circular harmonic trap, we prescribe the local change of the particle velocity due to the trapping force as
\begin{equation}
\vec{v}_{\mathrm{ext}}(\vec{r},t) =
\begin{cases}
-k(\vec{r} - \vec{r}_\mathrm{c}) , &\text{if } \norm{\vec{r} - \vec{r}_\mathrm{c}}<r_\mathrm{t},\\
0,  &\textrm{else} 
\end{cases}\label{eqn:HarmonicTrap}%
\end{equation}
with the parameter $k$ determining the trapping strength, the center of the trap $\vec{r}_\mathrm{c}$ being in the center of our simulation domain, and the radius of the trap $r_\mathrm{t}$. We varied $k r_\mathrm{t}/v_0\in[0,1]$ and $r_\mathrm{t}/\sigma\in[0,24]$ and used a quadratic simulation domain with edge length $256\sigma$ and periodic boundary conditions. 
The initial particle distribution is uniform and random with overall packing density $\Phi= \pi\bar{\rho} \sigma^2/4 = 0.2$, where $\bar{\rho}$ is the spatially averaged particle-number density, resulting in $\approx 17,000$ particles in the system. Since the simulation domain is much larger than the largest trap considered in this work and since the particles' packing density is moderate, the particle density outside the trap does not change for more than $\approx 10\%$ even if so many particles are trapped that the largest trap is closely packed with particles. 
We use Lennard-Jones units and choose the particle diameter $\sigma$, Lennard-Jones time $\tau_\mathrm{LJ} = \sigma^2/(\beta D_\mathrm{T} \varepsilon)$, and interaction energy $\varepsilon$ as units of length, time, and energy, respectively. 
The translational diffusion coefficient of a particle is given by $D_\mathrm{T} = \sigma^2/\tau_\mathrm{LJ}$ in Lennard-Jones units. 
We describe the activity of the particles by the dimensionless P\'eclet number $\mathrm{Pe} = v_0 \sigma/D_\mathrm{T}$, for which we choose $\mathrm{Pe} = 100$ in all simulations. In addition, we choose $v_0=24\sigma/\tau_\mathrm{LJ}$ to be consistent with previous works \cite{Stenhammar2014,stenhammar2015activity,Solon2015a,Stenhammar16,RW,Jeggle2019,BickmannBJW2020Chiraliry}.
The Langevin equations \eqref{eqn:r-LangevinAllgemein} and \eqref{eqn:phi-Langevin} are solved for a total simulation time of $2000 \tau_\mathrm{LJ}$ with a time-step size of $5 \cdot 10^{-5} \tau_\mathrm{LJ}$.

To evaluate the collective dynamics of the ABPs, we calculate the mean dimensionless interaction energy per particle $E_\mathrm{int}/(\varepsilon N_\mathrm{t})$ inside a trap, where $E_\mathrm{int}$ is the total interaction energy of the particles with a distance less than $r_\mathrm{t}$ from the center of the trap and $N_\mathrm{t}$ is their number. 
An advantage of using $E_\mathrm{int}/(\varepsilon N_\mathrm{t})$ as measure for the evaluation is that it allows to easily distinguish between a rather loose accumulation of the particles by the trapping force and MIPS.
To ensure reliable and robust results, we measure after an initial simulation period of $1600 \tau_\mathrm{LJ}$ the interaction energy per particle $10$ times with a period of $40 \tau_\mathrm{LJ}$ between each measurement and average over the individual measurements.

For some simulations, where strong traps were considered, the time-step size was halved. 
When passive particles, which move slowly, are combined with large and strong traps, which collect many particles until they are filled, the simulation time and the period between measurements were doubled. 
An overview about the different simulations and their time-step sizes, simulation times, and periods between measurements is provided by the Supplemental Material \cite{SI}.

\section{\label{sec:results}Results}
We consider two different external force fields: gravity and harmonic traps. All presented results are obtained for 2D and an activity of $\mathrm{Pe} = 100$.

\subsection{Gravity}
In the presence of gravity, the external-force contribution to the particle velocity reads $\vec{v}_{\mathrm{ext}} = -v_\mathrm{g} \vec{e}_x$, where $v_\mathrm{g}$ is the sedimentation speed caused by the gravitational field and $\vec{e}_x$ is the unit vector in $x$ direction. Equation \eqref{eqn:StationaryEq} now reduces to the one-dimensional differential equation
\begin{equation}
D(\rho)\frac{\partial}{\partial x}\rho = -v_\mathrm{g}\rho.
\label{eqn:GravityDGL}%
\end{equation}

In the dilute limit, where $D(\rho)\rightarrow D_\mathrm{T} + a_0$, this equation is solved by the Boltzmann distribution 
\begin{equation}
\rho_\mathrm{B}(x) = \rho_{\mathrm{B},0} e^{-\frac{v_\mathrm{g}x}{D_\mathrm{T} + a_0}}
\label{eqn:GravityDGLB}%
\end{equation}
with $\rho_{\mathrm{B},0} = \rho_\mathrm{B}(0)$. 
Here, $D_\mathrm{T}+a_0$ is the effective diffusion coefficient of the ABPs that corresponds to an effective temperature 
$T_{\mathrm{eff}} = (D_\mathrm{T}+a_0)\gamma/k_\mathrm{B} = T(1+v_0^2\tau_{\mathrm{LJ}}^2/(6\sigma^2))$ of the system, where $\gamma=k_\mathrm{B} T/D_\mathrm{T}$ denotes the translational friction coefficient of a particle. 
Since the coefficient $a_0$ is nonnegative, Eq.\ \eqref{eqn:GravityDGLB} suggests that in a gravitational field the density of ABPs decreases slower for increasing $x$ than the density of passive Brownian particles at the same temperature $T$. Note that not the physical temperature $T$ of the solvent surrounding the particles but the effective temperature $T_{\mathrm{eff}}$ that can be associated with the particle motion \cite{Palacci2010} describes the slope of the exponential decay. 

For higher densities, one needs to solve Eq.\ \eqref{eqn:GravityDGL} numerically, since $D(\rho)$, given by Eq.\ \eqref{eqn:DDdiffusion}, is now density-dependent. Obtaining the corresponding numerical solution $\rho_\mathrm{I}(x)$, however, requires knowledge of the values of the coefficients $\lbrace a_i\rbrace$ and of the velocity $v_\mathrm{g}$. To determine the values of $\lbrace a_i\rbrace$ for the considered ABPs, we use Eqs.\ \eqref{eqn:2Da_0}-\eqref{eqn:2Da_2} and \eqref{eq:A100}-\eqref{eq:A01m1} from Appendix \ref{app:A}. Furthermore, we choose the gravity-induced velocity $v_\mathrm{g}=v_{0}/24$. 
To make the two cases with and without interactions of the ABPs comparable, we choose $\rho_{\mathrm{I}}(0) = 8/(5\pi\sigma^2)$ as boundary condition for the interacting particles and enforce equal particle numbers in both cases by the condition $\int_0^\infty\!\rho_{\mathrm{B}}\,\dif x = \int_0^\infty\!\rho_{\mathrm{I}}\,\dif x$. Note that the boundary condition for $\rho_{\mathrm{I}}$ implies a packing density $\rho_{\mathrm{I}}(0)\pi\sigma^2/4 = 0.4$ of the particles at $x=0$, which is close to the critical packing density of $\approx 0.434$ \cite{BickmannW2020} that would lead to MIPS.
\begin{figure}[htpb]
\centering
\includegraphics[width = \linewidth]{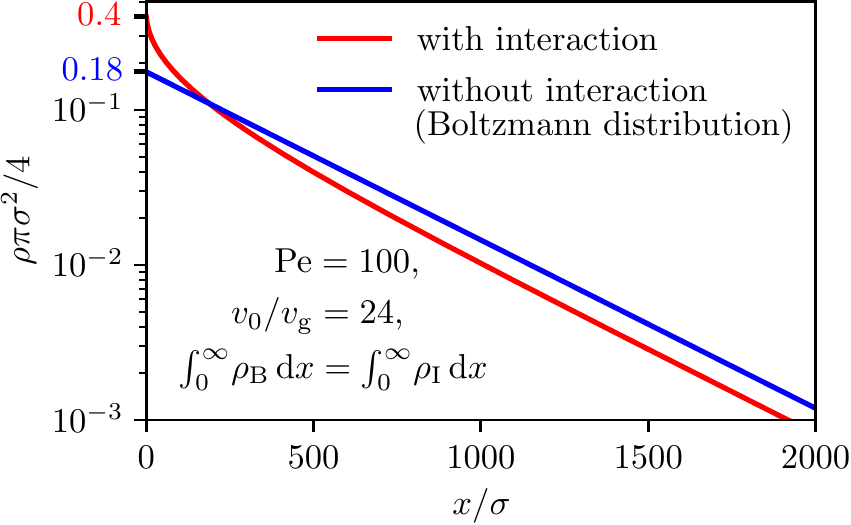}
\caption{\label{fig:1}Numerical solution of Eq.\ \eqref{eqn:GravityDGL} for interacting ABPs under gravity (red) and Boltzmann distribution \eqref{eqn:GravityDGLB} for noninteracting ABPs under gravity (blue) for $v_0/v_\mathrm{g} = 24$. We have chosen $\rho_{\mathrm{I}}(0)\pi\sigma^2/4 = 0.4$ for the curve representing the interacting particles. The two functions are normalized so that the total number of particles of the system is the same for both cases. This leads to $\rho_{\mathrm{B}}(0)\pi\sigma^2/4 =\rho_{\mathrm{B},0}\pi\sigma^2/4 \approx 0.18$. See Supplemental Material \cite{SI} for the raw data corresponding to this figure.}
\end{figure}

The results are shown in Fig.\ \ref{fig:1}. We see deviations between the curves for interacting and noninteracting ABPs for small values of $x$, i.e., in the high-density regime where the density dependence of $D(\rho)$ is important: The curve that considers interactions does not follow a Boltzmann distribution. With interactions the density of the ABPs is larger than without interactions in this regime. 
This originates from the fact that the activity creates an effective attractive interaction potential \cite{farage2015effective} that replaces the purely repulsive interaction of the particles \cite{MariniBettoloMarconi2015} and leads to accumulation where interactions are relevant. 
The effective attractive interaction enters Eq.\ \eqref{eqn:GravityDGL} through the coefficients $a_1$ and $a_2$ in the density-dependent diffusion coefficient \eqref{eqn:DDdiffusion} and their contributions increase with $\rho$. For large values of $v_0$, as they are considered here, the contribution of the coefficient $a_1$ reduces the value of $D(\rho)$, which leads to the observed accumulation of ABPs. 
Interestingly, our analytic approach reveals that this mechanism is the same as that leading to MIPS for even larger densities. 
To see this, one has to take into account that the coefficient $a_1$ determines also the density-dependence of the mean swimming speed of ABPs \cite{BickmannW2020,BickmannW2020b} and that a sufficiently strong decrease of the mean swimming speed with increasing density leads to MIPS \cite{MIPS}. 
For large values of $x$, i.e., in the low-density regime, both curves show the same qualitative behavior, since interactions between particles are rare in this regime. The curves now follow an exponential decay with the same decay constant $-v_\mathrm{g}/(D_\mathrm{T}+a_0)$. Due to particle-number conservation, however, the density of interacting ABPs is smaller than the density of noninteracting ABPs for large values of $x$.

\subsection{Harmonic traps}
\begin{figure*}[htpb]
\includegraphics[width=\linewidth]{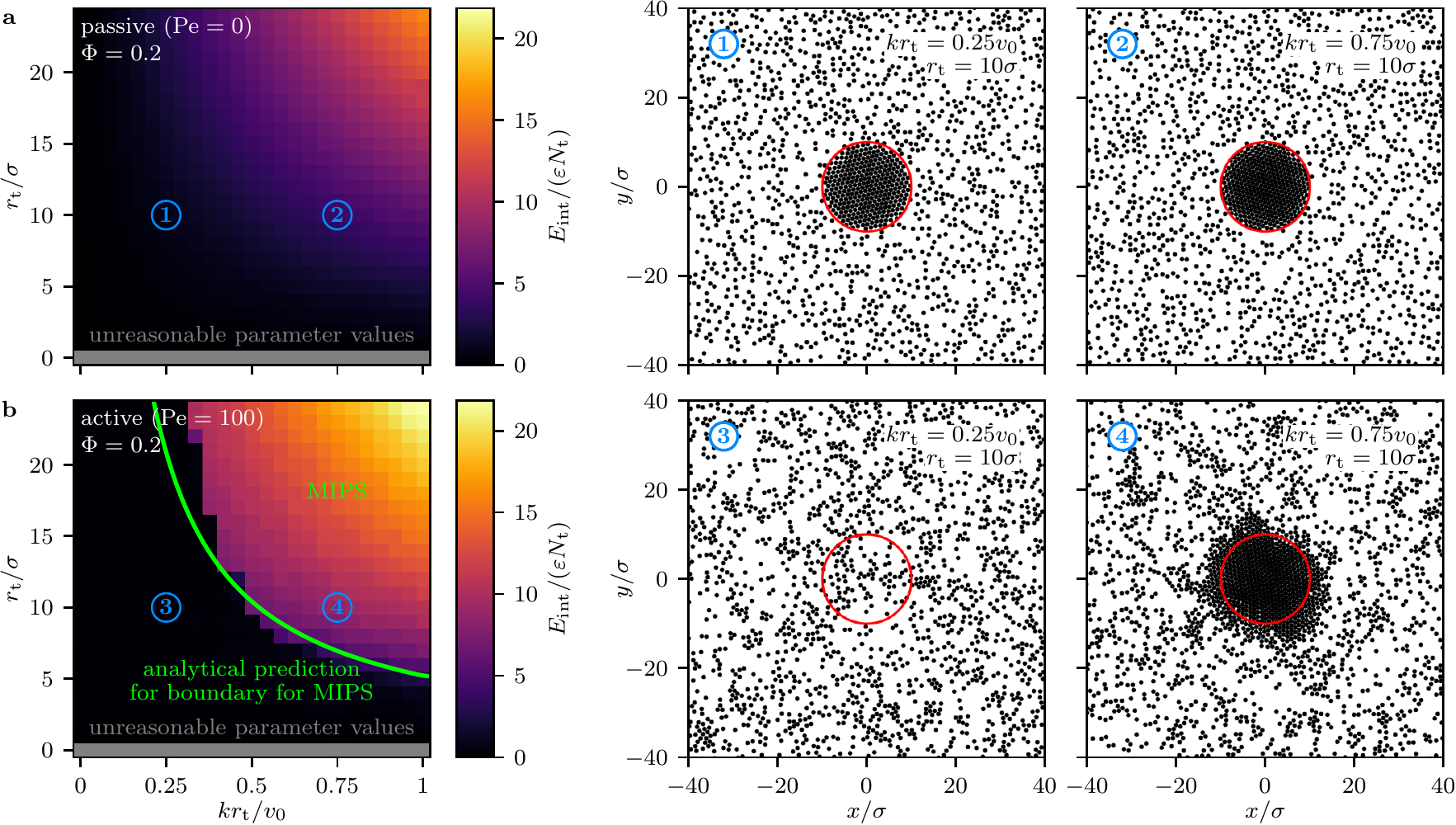}
\caption{\label{fig:fig2}Mean dimensionless interaction energy per particle $E_\mathrm{int}/(\varepsilon N_\mathrm{t})$ inside a harmonic trap as a function of trapping strength $k$ and trap radius $r_\mathrm{t}$. 
(a) In the case of passive particles, $E_\mathrm{int}/(\varepsilon N_\mathrm{t})$ increases smoothly when $k$ or $r_\mathrm{t}$ are increased. Respective snapshots show that passive particles always accumulate inside the trap. (b) In contrast, for active particles $E_\mathrm{int}/(\varepsilon N_\mathrm{t})$ increases suddenly and sharply when entering a region of sufficiently large $k$ and $r_\mathrm{t}$ where the ABPs undergo MIPS. Respective snapshots show that ABPs do not accumulate in a weak trap as their self-propulsion allows them to escape, whereas a strong trap allows for a MIPS cluster emerging inside the trap. Since ABPs accumulate at boundaries, the cluster formed in a trap can extend beyond the trap's boundary. 
See Supplemental Material \cite{SI} for the raw data corresponding to this figure.}
\end{figure*}
We consider circular harmonic traps of the form \eqref{eqn:HarmonicTrap} with adjustable trapping strength $k$ and trap radius $r_\mathrm{t}$.
This form of a trap is popular in numerical studies on the collective behavior of ABPs in external force fields \cite{yang2017artificial,yang2018colloidal,shen2019spatial,ribeiro2020trapping}. In experiments, such harmonic traps have already been realized by using optic \cite{xu2018optical} and acoustic \cite{lee2010transverse,Takatori2016} tweezers. 
Passive Brownian particles in such harmonic traps accumulate in the center. ABPs, on the other hand, can show many more effects: Their self-propulsion can overcome an attractive trap potential \cite{yang2017artificial}, they can confine passive particles inside a trap \cite{yang2018colloidal}, and they can form active shells \cite{yang2017artificial,sandoval2018self}. It has also been shown experimentally that active particles inside traps can be found in the center or at a certain distance from the center, depending on the propulsion speed of the particles and strength of the trap \cite{shen2019spatial}.

Here, we focus on the accumulation of particles and the occurrence of MIPS in the traps. 
By Brownian dynamics simulations, we calculate the mean dimensionless interaction energy per particle $E_\mathrm{int}/(\varepsilon N_\mathrm{t})$ \cite{BickmannBJW2020Chiraliry} inside the trap for various trapping strengths $k$ and trap radii $r_\mathrm{t}$. 
As initial condition, we use a homogeneous packing density of $\Phi = 0.2$. For comparison, we consider also the phase behavior of corresponding passive particles ($v_0 = 0$). 
The occurrence of MIPS inside a harmonic trap requires that the density of active particles in the trap is sufficiently high. We predict the onset of MIPS by numerically determining when $D(\rho)$ in Eq.\ \eqref{eqn:StationaryEq} becomes zero. The results are shown in Fig.\ \ref{fig:fig2}. 

Due to the trapping force, passive particles always accumulate inside the trap (see Fig.\ \ref{fig:fig2}a). For a stronger or larger trap, the packing density of passive particles in the center of the trap is larger (see snapshots 1 and 2 in Fig.\ \ref{fig:fig2}a). Therefore, the mean dimensionless interaction energy per particle $E_\mathrm{int}/(\varepsilon N_\mathrm{t})$ increases smoothly with the trapping strength $k$ and trap radius $r_\mathrm{t}$ (see state diagram in Fig.\ \ref{fig:fig2}a). 
 
ABPs, on the other hand, behave quite differently (see Fig.\ \ref{fig:fig2}b). Due to their self-propulsion, they can escape from weak or small traps, resulting in very small values of $E_\mathrm{int}/(\varepsilon N_\mathrm{t})$ for small $k$ or $r_\mathrm{t}$. An example is a trap with parameters $kr_{\mathrm{t}}=0.25v_0$ and $r_{\mathrm{t}}=10 \sigma$ (see snapshot 3 in Fig.\ \ref{fig:fig2}b). In this case, no accumulation of ABPs is found albeit passive particles would accumulate inside the same trap (cf.\ snapshot 1 in Fig.\ \ref{fig:fig2}a). When the trap becomes stronger and larger, the ABPs can remain inside the trap for a longer time and thus enhance the local concentration of ABPs. If the local density inside the trap increases so far that it becomes larger than the critical density for the onset of MIPS \cite{MIPS,RW,BickmannW2020}, particle clustering emerges. Note that we chose the activity of the particles sufficiently large to allow for MIPS \cite{Jeggle2019}. As MIPS clusters are tightly packed, $E_\mathrm{int}/(\varepsilon N_\mathrm{t})$ suddenly increases when MIPS occurs (see state diagram in Fig.\ \ref{fig:fig2}b). 
 
Our analytical prediction for the onset of MIPS and the sudden increase of $E_\mathrm{int}/(\varepsilon N_\mathrm{t})$ in the simulation data are in very good agreement.
They deviate only slightly for very large traps. We believe that this discrepancy originates from the fact that cluster formation inside a large trap significantly reduces the overall packing density around the trap, whereas our analytical approach assumes that density to be constant at $\Phi = 0.2$. 
Furthermore, the increase of $E_\mathrm{int}/(\varepsilon N_\mathrm{t})$ that is associated with entering the MIPS region in the state diagram is less pronounced for very small but strong traps. 
This can be explained as follows: For a very small trap, the overall number of ABPs inside the trap is very small, which makes $E_\mathrm{int}/(\varepsilon N_\mathrm{t})$ quite susceptible to Brownian fluctuations. 
While the accumulation of passive particles is restricted to the trap, ABPs can form clusters that extend beyond the boundary of the trap. This effect can be observed, e.g., for a trap with parameters $kr_{\mathrm{t}}=0.75v_0$ and $r_{\mathrm{t}}=10 \sigma$ (see snapshot 4 in Fig.\ \ref{fig:fig2}b), which corresponds to a point that is well inside the MIPS region in the state diagram.  
Moreover, the effect is closely related to the known phenomenon that ABPs accumulate at walls or other stationary obstacles \cite{vladescu2014filling,yang2014}, where the MIPS cluster inside the trap here acts as a fixed obstacle.

\section{\label{sec:Conclusion}Conclusions}
We investigated the collective dynamics of ABPs under the influence of external force fields in 2D and 3D. For this purpose, we derived predictive field theories from the Langevin equations that describe the motion of the ABPs on a microscopic level. These field theories are applicable for small to large densities and activities of the particles. In particular, the derived field theories are an advection-diffusion model, which contains up to two spatial derivatives per term and is the main focus of our current investigation, and a phase-field model, which contains derivatives up to fourth order. 
With the advection-diffusion model, we studied the effect of gravitation and harmonic traps on the steady state of the ABPs. For ABPs under gravity, we obtained a modified Boltzmann distribution that takes the interactions and activity of the particles into account. 
In the case of the harmonic traps, we predicted for which trap sizes and trapping forces the traps induce MIPS in the system. To confirm our predictions, we performed Brownian dynamics simulations. Their results were found to be in very good agreement with our analytical predictions. 
In summary, our results show that ABPs in external fields can exhibit interesting effects that arise from the coupling of the external force, the interactions of the particles, and their activity. 
An understanding of this coupling is helpful for future applications of ABPs \cite{yang2018colloidal,nitschkeSW2021collective}, where external fields are likely to be relevant, and available through our field theories. 

As, according to our results, the occurrence of MIPS can be controlled quite arbitrarily by external fields, such fields allow to realize programmable density patterns in systems containing ABPs. Since active particles are known to have a strong effect on the behavior of passive particles \cite{stenhammar2015activity}, it is likely that also programmable materials that contain both active and passive particles can be realized.
In the future, one should continue this study towards time-dependent external fields, for which our models are applicable as well if the fields do not change too fast with time. 
When extending our phase-field model by including terms up to sixth order in derivatives, one could use the extended model to study ABPs in high-gravity regions that can induce crystallization of the ABPs \cite{loffler2002crystallization,he2019high}.
Furthermore, one could take external torques into account, which can result in a nematic ordering of the particles allowing for interesting effects like floating phases \cite{schmidt2004floating,de2012floating}.

\begin{acknowledgments}
We thank Simon Hartmann and Uwe Thiele for helpful discussions.
R.W.\ is funded by the Deutsche Forschungsgemeinschaft (DFG, German Research Foundation) -- WI 4170/3-1. 
The simulations for this work were performed on the computer cluster PALMA II of the University of M\"unster.
\end{acknowledgments}

\appendix
\section{\label{app:A}Microscopic expressions for the coefficients occurring in Eqs.\ (\ref{eqn:DDdiffusion}) and (\ref{eqn:4thOrderEXTcurrent})}
The coefficients $\lbrace a_i\rbrace$ in Eq.\ \eqref{eqn:DDdiffusion} and $\lbrace b_i\rbrace$ in Eq.\ \eqref{eqn:4thOrderEXTcurrent} can be related to microscopic properties of the system. In the following, we present the corresponding expressions for 2D and 3D.

\subsection{\label{app:2DA}Two spatial dimensions}
In 2D, the coefficients $\lbrace a_i\rbrace$ and $\lbrace b_i\rbrace$ are given by 
\begin{align}
a_0 &= \frac{v_0^2}{2D_\mathrm{R}},\label{eqn:2Da_0}\\
a_1 &= \frac{A(1,0,0)}{\pi} - \frac{v_0}{\pi D_\mathrm{R}}(2A(0, 1, 0) + A(0, 1, -1)),\\
a_2 &= \frac{4}{\pi^2 D_\mathrm{R}}A(0,1,0)(A(0,1,0)+A(0,1,-1)),\label{eqn:2Da_2}\\
b_1 &=  \frac{v_0^2}{2D_\mathrm{R}^2},\\
b_2 &= -\frac{v_0}{\pi D_\mathrm{R}^2}(A(0,1,-1)+3A(0,1,0)),\\
b_3 &= \frac{4 A(0,1,0)}{\pi^2 D_\mathrm{R}^2}(A(0,1,-1)+A(0,1,0)),\\
b_4 &= -\frac{2v_0 A(0,1,0)}{\pi D_\mathrm{R}^2}.\label{eqn:2Db_4}
\end{align}
We follow the notation of Ref.\ \cite{BickmannW2020}, which gives approximate expressions for the coefficients $A(n, k_1, k_2)$ that originate from the pair-distribution function of the particles given in Ref.\ \cite{Jeggle2019}. The coefficients occurring in Eqs.\ \eqref{eqn:2Da_0}-\eqref{eqn:2Db_4} are approximately given by \cite{BickmannW2020,Jeggle2019}
\begin{align}
A(1, 0, 0) &= (38.2 + 18.4e^{2.87\Phi})\sigma^4 \tau_{\mathrm{LJ}}^{-1},\label{eq:A100}\\
A(0, 1,0) &= 36.95\sigma^3 \tau_{\mathrm{LJ}}^{-1},\\
A(0, 1,-1) &= -(0.23 + 13.6 \Phi)\sigma^3 \tau_{\mathrm{LJ}}^{-1}.\label{eq:A01m1}
\end{align}

\subsection{\label{app:3DA}Three spatial dimensions}
In 3D, the coefficients $\lbrace a_i\rbrace$ and $\lbrace b_i\rbrace$ are given by
\begin{align}
a_0 &= \frac{v_0^2}{6D_\mathrm{R}},\label{eqn:3Da_0}\\
\begin{split}
a_1 &= \frac{2G(1,0,0,0)}{3\pi}\\
&\quad\,\, - \frac{v_0}{3\pi D_\mathrm{R}}(3G(0, 1, 1, 0) + G(0, 1, 0, 1)),
\end{split}\\
a_2 &= \frac{4}{3\pi^2 D_\mathrm{R}}G(0,1,1,0)(G(0,1,1,0)+G(0,1,0,1)),\\
b_1 &= \frac{v_0^2}{12 D_\mathrm{R}^2},\\
b_2 &= -\frac{v_0}{6 \pi D_\mathrm{R}^2}(G(0,1,0,1)+3G(0,1,1,0)),\\
b_3 &= \frac{2 G(0,1,1,0)}{3 \pi^2 D_\mathrm{R}^2}(G(0,1,0,1)+G(0,1,1,0)),\\
b_4 &= -\frac{v_0 G(0,1,1,0)}{3 \pi D_\mathrm{R}^2}.\label{eqn:3Db_4}
\end{align}
Here, we use the notation of Ref.\ \cite{BickmannW2020b} for the coefficients $G(n, l_1, l_2, l_3)$ originating from the pair-distribution function given in Ref.\ \cite{Broeker2019}. The coefficients occurring in Eqs.\ \eqref{eqn:3Da_0}-\eqref{eqn:3Db_4} are approximately given by 
\begin{align}
G(1,0,0,0) &= (41.59 + 12.69 e^{4.07\Phi})\sigma^5 \tau_{\mathrm{LJ}}^{-1},\\
G(0, 1, 1, 0) &= (22.49 + 7.05\Phi)\sigma^4 \tau_{\mathrm{LJ}}^{-1},\\
G(0,1,0,1) &= -20.48\sigma^4 \tau_{\mathrm{LJ}}^{-1},
\end{align}
where $\Phi$ denotes the overall packing density in 3D that is related to the spatially averaged particle-number density $\bar{\rho}$ by $\Phi = \pi\bar{\rho}\sigma^3/6$ \cite{BickmannW2020b,Broeker2019}.

\section{\label{app:Jint}Density current $\boldsymbol{\vec{J}^{(\mathrm{int})}}$}
The density current $\vec{J}^{(\mathrm{int})}$ in Eq.\ \eqref{eq:Modell4O} is given by \cite{BickmannW2020,BickmannW2020b}
\begin{align}
J_{\mathrm{int}, i} &= D(\rho)\partial_i\rho \nonumber\\
&\quad +(\alpha_{3}+\alpha_{4}\rho+\alpha_{5}\rho^2+\alpha_{6}\rho^3+\alpha_{7}\rho^4)\partial_i\Laplace\rho\nonumber\\
&\quad +(\alpha_{8}+\alpha_{9}\rho+\alpha_{10}\rho^2+\alpha_{11}\rho^3)(\Laplace\rho)\partial_i\rho \label{eqn:4thorderCurrentJint}\\
&\quad +(\alpha_{12}+\alpha_{13}\rho+\alpha_{14}\rho^2+\alpha_{15}\rho^3)(\partial_j\rho)\partial_i\partial_j\rho\nonumber\\
&\quad +(\alpha_{16}+\alpha_{17}\rho+\alpha_{18}\rho^2)(\partial_j\rho)(\partial_j\rho)\partial_i\rho,\nonumber
\end{align}
where the coefficients $\lbrace a_i\rbrace$ depend on the dimensionality of the system. 
For two spatial dimensions, they read \cite{BickmannW2020}
\begin{align}
\begin{split}
\alpha_{3} &= \frac{\tau^2 v_0^2}{32} (16 D_\mathrm{T}+\tau  v_0^2),
\end{split}
\\
\alpha_{4} &= \frac{1}{16\pi}(-16 D_\mathrm{T} \tau ^2 v_0 (A(0,1,-1)+3 A(0,1,0))\nonumber\\
&\quad\, +\tau  v_0 (\tau  v_0 (-\tau  v_0 (A(0,1,-1) +5 A(0,1,0)\nonumber\\
&\quad\, +A(0,1,1))+4 A(1,0,-1)+4 A(1,0,1)\\
&\quad\, +A(1,2,-2)+4 A(1,2,-1) +A(1,2,0))\nonumber\\
&\quad\, -6A(2,1,-1)-6A(2,1,0))+2 A(3,0,0)),\nonumber
\\
\alpha_{5} &= \frac{\tau}{8\pi^2} (32 D_\mathrm{T} \tau  A(0,1,0) (A(0,1,-1)+A(0,1,0))\nonumber\\
&\quad\, +\tau ^2 v_0^2 (A(0,1,-1) (4 A(0,1,0)+A(0,1,1))\nonumber\\
&\quad\, +A(0,1,0) (9 A(0,1,0)+4 A(0,1,1)))\nonumber\\
&\quad\, -\tau  v_0(A(0,1,1) A(1,2,-2)+A(0,1,-1) \nonumber\\
&\qquad\;\, (4A(1,0,-1)+4A(1,0,1)+4A(1,2,-1)\nonumber\\
&\quad\,+A(1,2,0))+A(0,1,0) (3 (4 A(1,0,-1)\\
&\quad\, +4 A(1,0,1)+A(1,2,-2)+4 A(1,2,-1))\nonumber\\
&\quad\, +2 A(1,2,0)) ) +A(1,2,-2) A(1,2,0)\nonumber\\
&\quad\,+12 A(0,1,0) A(2,1,-1)+6 (A(0,1,-1)\nonumber\\
&\quad\, +A(0,1,0)) A(2,1,0)), \nonumber
\\
\alpha_{6} &= -\frac{\tau^2}{4\pi^3}  A(0,1,0)(\tau  v_0 (A(0,1,-1) (5 A(0,1,0)\nonumber\\
&\quad\, +3 A(0,1,1))+A(0,1,0) (7 A(0,1,0)\nonumber\\
&\quad\,\quad\, (A(1,0,-1)+5 A(0,1,1))-8A(0,1,-1)\nonumber\\
&\quad\,-8A(0,1,0)+A(1,0,1))-2A(1,2,-2)\\
&\quad\,\quad\,(A(0,1,0)+A(0,1,1))-8A(1,2,-1)\nonumber\\
&\quad\,\quad\,(A(0,1,-1)+A(0,1,0)) -A(1,2,0)\nonumber\\
&\quad\,\quad\,(A(0,1,-1)+A(0,1,0)) )),\nonumber 
\\
\begin{split}
\alpha_{7} &= \frac{\tau^3}{\pi^4} A(0,1,0)^2 (A(0,1,-1)+A(0,1,0))\\
&\quad\,\;(A(0,1,0)+A(0,1,1)),
\end{split}
\\
\alpha_{8} &= -\frac{\tau v_0}{8\pi} (16 D_\mathrm{T} \tau  A(0,1,0)+2A(2,1,0)\nonumber\\
&\quad\,+\tau ^2 v_0^2 (4 A(0,1,-1)+A(0,1,0))\\
&\quad\,-4 \tau  v_0 (A(1,0,0)+A(1,2,-1))),\nonumber
\\
\alpha_{9} &= \frac{\tau}{8\pi^2} (4 A(0,1,0) (8 D_\mathrm{T} \tau  (A(0,1,-1)+A(0,1,0))\nonumber\\
&\quad\,+3 A(2,1,-1))+\tau ^2 v_0^2(8 A(0,1,-1)^2\nonumber\\
&\quad\,+A(0,1,-1)(42 A(0,1,0)+A(0,1,1))\nonumber\\
&\quad\,+6 A(0,1,0)^2-A(0,1,1)^2)\nonumber\\
&\quad\,+\tau v_0(-A(0,1,1) (A(1,2,-2)-2 A(1,2,0))\nonumber\\
&\quad\,-A(0,1,0) (8 A(1,0,-1)+24 A(1,0,0)\\
&\quad\,+8 A(1,0,1)+3 A(1,2,-2)+36 A(1,2,-1)\nonumber\\
&\quad\,-2 A(1,2,0))-A(0,1,-1) (8 A(1,0,0)\nonumber\\
&\quad\,+4 A(1,2,-1)+A(1,2,0)))+A(1,2,0)\nonumber\\
&\quad\,\quad\,(A(1,2,-2)-A(1,2,0)) +2 A(2,1,0)\nonumber\\
&\quad\,\quad\,(A(0,1,-1)+3 A(0,1,0)) ),  \nonumber
\\
\alpha_{10} &= - \frac{\tau^2 }{2\pi^3} A(0,1,0)(\tau  v_0 (16 A(0,1,-1)^2\nonumber\\
&\quad\, +2 A(0,1,-1)(17 A(0,1,0)+A(0,1,1)) \nonumber\\
&\quad\,+(A(0,1,0)-A(0,1,1)) (3 A(0,1,0)\nonumber\\
&\quad\,+2 A(0,1,1)))-4 (A(0,1,-1)+A(0,1,0))\nonumber\\
&\qquad\,\,(A(1,0,-1)+2 A(1,0,0)+A(1,0,1))\\
&\quad\,-2 A(1,2,-2)(A(0,1,0)+A(0,1,1)) \nonumber\\
&\quad\,-8 A(1,2,-1)(A(0,1,-1)+2 A(0,1,0)) \nonumber\\
&\quad\,+A(1,2,0)(A(0,1,0)+2 A(0,1,1)) ), \nonumber
\\
\alpha_{11} &= \frac{\tau^3}{\pi^4} A(0,1,0)^2 (16 A(0,1,-1)^2+A(0,1,-1)\nonumber\\
&\quad\,\quad\,(A(0,1,1)+17 A(0,1,0)) +(A(0,1,0)\\
&\quad\,-2 A(0,1,1))(A(0,1,0)+A(0,1,1))),\nonumber
\\
\alpha_{12} &= -\frac{\tau v_0}{4\pi} (16 D_\mathrm{T} \tau A(0,1,0)+2A(2,1,0)\nonumber\\
&\quad\,+\tau ^2 v_0^2 (2 A(0,1,-1)+2 A(0,1,0)\nonumber\\
&\quad\,+3 A(0,1,1))-\tau  v_0 (2 A(1,0,-1)\\
&\quad\, +2 A(1,0,0)+2 A(1,0,1)+3 A(1,2,0))),\nonumber
\\
\alpha_{13} &= \frac{\tau}{4\pi^2}(32 D_\mathrm{T} \tau  A(0,1,0) (A(0,1,-1)\nonumber\\
&\quad\,+A(0,1,0))+\tau ^2 v_0^2(4 A(0,1,-1)^2 \nonumber\\
&\quad\,+A(0,1,-1)(24 A(0,1,0)+5 A(0,1,1)) \nonumber\\
&\quad\,+14 A(0,1,0)^2+32 A(0,1,0) A(0,1,1)) \nonumber\\
&\quad\,+A(0,1,1)^2-\tau  v_0(A(0,1,1) (A(1,2,-2)\nonumber\\
&\quad\,+2 A(1,2,0))+A(0,1,-1) (2 A(1,0,-1)\nonumber\\
&\quad\,+4 A(1,0,0)+2 A(1,0,1)+5 A(1,2,0))\\
&\quad\,+A(0,1,0) (22 A(1,0,-1)+8 A(1,2,-1)\nonumber\\
&\quad\,+12 A(1,0,0)+22 A(1,0,1)+3 A(1,2,-2)\nonumber\\
&\quad\,+18 A(1,2,0)))+A(1,2,-2) A(1,2,0)\nonumber\\
&\quad\,+A(1,2,0)^2+12 A(0,1,0) A(2,1,-1)\nonumber\\
&\quad\,+2 A(0,1,-1) A(2,1,0) +6 A(0,1,0) A(2,1,0)),\nonumber
\\
\alpha_{14} &= -\frac{\tau^2 }{\pi^3} A(0,1,0)(\tau  v_0 (8 A(0,1,-1)^2\nonumber\\
&\quad\,+A(0,1,-1)(21 A(0,1,0)+11 A(0,1,1)) \nonumber\\
&\quad\,+8 A(0,1,0)^2+26 A(0,1,1) A(0,1,0)\nonumber\\
&\quad\,+2A(0,1,1)^2)-A(0,1,-1) (6 A(1,0,-1)\nonumber\\
&\quad\,+4 A(1,0,0)+6 A(1,0,1)+4 A(1,2,-1)\\
&\quad\,+5 A(1,2,0))-2 (A(0,1,1) (A(1,2,-2)\nonumber\\
&\quad\,+A(1,2,0))+A(0,1,0) (5 A(1,0,-1)\nonumber\\
&\quad\,+2 A(1,0,0)+5 A(1,0,1)+A(1,2,-2)\nonumber\\
&\quad\,+2 A(1,2,-1)+3 A(1,2,0)))), \nonumber
\\
\alpha_{15} &= \frac{2}{\pi^4} \tau ^3 A(0,1,0)^2(8 A(0,1,-1)^2 +11 A(0,1,-1)\nonumber\\
&\quad\,\quad\,(A(0,1,0)+A(0,1,1)) +3 A(0,1,0)^2\\
&\quad\,+2 A(0,1,1)^2+13 A(0,1,0) A(0,1,1)),\nonumber
\\
\alpha_{16} &= \frac{\tau}{8\pi^2} (\tau ^2 v_0^2 (24 A(0,1,1) A(0,1,0)+A(0,1,1)^2\nonumber\\
&\quad\,+7 A(0,1,0)^2+32 A(0,1,-1) A(0,1,0))\nonumber\\
&\quad\,-2 \tau  v_0(A(0,1,1) A(1,2,0)+A(0,1,0)\\
&\qquad\,\, (8 A(1,0,-1)+8 A(1,0,0)+8 A(1,0,1)\nonumber\\
&\quad\,+8 A(1,2,-1)+6 A(1,2,0))) +A(1,2,0)^2),  \nonumber
\\
\alpha_{17} &= \frac{\tau^2}{4\pi^3} A(0,1,0)(-\tau  v_0 (32 A(0,1,-1)^2\nonumber\\
&\quad\,+A(0,1,-1)(103 A(0,1,0)+21 A(0,1,1)) \nonumber\\
&\quad\,+15 A(0,1,0)^2+71 A(0,1,0) A(0,1,1))\nonumber\\
&\quad\,+6 A(0,1,1)^2+4A(0,1,1) A(1,2,-2)\nonumber\\
&\quad\,+8 A(0,1,-1) (A(1,0,-1)+2 A(1,0,0)\\
&\quad\,+A(1,0,1)+A(1,2,-1))+4A(0,1,0) \nonumber\\
&\quad\,\quad\,(6 A(1,0,-1)+4 A(1,0,0)+6 A(1,0,1)\nonumber\\
&\quad\,+A(1,2,-2)+6 A(1,2,-1))+A(1,2,0)\nonumber\\
&\quad\,\quad\,(6 A(0,1,1)+13 A(0,1,0)+9 A(0,1,-1)) ), \nonumber
\\
\alpha_{18} &= \frac{4}{\pi^4} \tau ^3 A(0,1,0)^2(8 A(0,1,-1)^2+A(0,1,0)^2\nonumber\\
&\quad\, +A(0,1,1)^2+A(0,1,-1)(9 A(0,1,0)\\
&\quad\, +5 A(0,1,1)) +6 A(0,1,0) A(0,1,1)),\nonumber
\end{align}
and for three spatial dimensions, they are \cite{BickmannW2020b}
\begin{align}
\alpha_{3} &= \frac{1}{540} \tau ^2 v_0^2 (45 D_\mathrm{T}+2 \tau  v_0^2),\\
    \alpha_{4} &= \frac{1}{270 \pi}(\tau  v_0(-45 D_\mathrm{T} \tau  (G(0,1,0,1)\nonumber\\
    &\quad\,+3 G(0,1,1,0))+\tau  v_0(\tau  v_0 (-2 G(0,1,0,1)\nonumber\\
    &\quad\,-10 G(0,1,1,0)+\sqrt{2} (G(0,1,1,2)\nonumber\\
    &\quad\,+G(0,1,2,1)))-5 \sqrt{3} G(1,0,1,1)\\
    &\quad\,+4 G(1,2,0,2)+2 \sqrt{30} G(1,2,1,1)\nonumber\\
    &\quad\,+4 G(1,2,2,0))-27 (G(2,1,0,1)\nonumber\\
    &\quad\,+G(2,1,1,0)))+18 G(3,0,0,0)),\nonumber
\\
    \alpha_{5} &= \frac{1}{135\pi^2} \tau (90 D_\mathrm{T} \tau  G(0,1,1,0) (G(0,1,0,1)\nonumber\\
    &\quad\,+G(0,1,1,0))+\tau ^2 v_0^2(18 G(0,1,1,0)^2\nonumber\\
    &\quad\,+G(0,1,1,2) G(0,1,2,1)-4 \sqrt{2} (G(0,1,1,2)\nonumber\\
    &\quad\,+G(0,1,2,1)) G(0,1,1,0)+G(0,1,0,1) \nonumber\\
    &\quad\quad\,\, (8 G(0,1,1,0)-\sqrt{2} (G(0,1,1,2)\nonumber\\
    &\quad\,+G(0,1,2,1))))+\tau  v_0 (G(0,1,1,0) \nonumber\\
    &\quad\,\quad\,(15 \sqrt{3} G(1,0,1,1)-12 G(1,2,0,2)\nonumber\\
    &\quad\,-6 \sqrt{30} G(1,2,1,1)-8 G(1,2,2,0))\\
    &\quad\,+G(0,1,0,1) (5 \sqrt{3} G(1,0,1,1)\nonumber\\
    &\quad\,-2 \sqrt{30} G(1,2,1,1)-4 G(1,2,2,0))\nonumber\\
    &\quad\, +2 \sqrt{2} (G(0,1,2,1) G(1,2,0,2)\nonumber\\
    &\quad\,+G(0,1,1,2) G(1,2,2,0)))+8 G(1,2,0,2) \nonumber\\
    &\quad\,\quad\,G(1,2,2,0)+54 G(0,1,1,0) G(2,1,0,1)\nonumber\\
    &\quad\,+27 (G(0,1,0,1)+G(0,1,1,0)) G(2,1,1,0)),\nonumber
\\
    \alpha_{6} &= \frac{2}{135\pi^3} \tau ^2 (-14 \tau  v_0 G(0,1,1,0)^3\nonumber\\
    &\quad\,+G(0,1,1,0)^2(5 \tau  v_0 (\sqrt{2}G(0,1,1,2)\nonumber\\
    &\quad\,+\sqrt{2}G(0,1,2,1)-2 G(0,1,0,1))\nonumber\\
    &\quad\,-10 \sqrt{3} G(1,0,1,1)+8G(1,2,0,2)\nonumber\\
    &\quad\,+4\sqrt{30}G(1,2,1,1)+4G(1,2,2,0)) \nonumber\\
    &\quad\,+G(0,1,1,0)( 3 \tau  v_0(\sqrt{2} G(0,1,0,1) \nonumber\\
    &\quad\,\quad\,(G(0,1,1,2)+G(0,1,2,1)) \nonumber\\
    &\quad\, -G(0,1,1,2) G(0,1,2,1))\\
    &\quad\,-2 \sqrt{2} (2 G(0,1,2,1) G(1,2,0,2)\nonumber\\
    &\quad\,+G(0,1,1,2) G(1,2,2,0))\nonumber\\
    &\quad\,+G(0,1,0,1) (4\sqrt{30}G(1,2,1,1)\nonumber\\
    &\quad\,+4G(1,2,2,0)-10 \sqrt{3} G(1,0,1,1)))\nonumber\\
    &\quad\, -G(0,1,0,1) G(0,1,1,2)(\tau  v_0 G(0,1,2,1) \nonumber\\
    &\quad\, +2 \sqrt{2} G(1,2,2,0))),\nonumber
\\
    \alpha_{7} &= \frac{8}{135\pi^4} \tau ^3 G(0,1,1,0) (G(0,1,0,1)\nonumber\\
    &\quad\,+G(0,1,1,0))(-\sqrt{2}G(0,1,1,0)\nonumber\\
    &\quad\,\quad\,(G(0,1,1,2)+G(0,1,2,1)) \\
    &\quad\,+2 G(0,1,1,0)^2 +G(0,1,1,2) G(0,1,2,1)), \nonumber
\\
    \alpha_{8} &= -\frac{1}{1080\pi} \tau v_0 (20 \tau (18 D_\mathrm{T}+\tau v_0^2) G(0,1,1,0)\nonumber\\
    &\quad\,+\tau  v_0 (\tau v_0 (30 G(0,1,0,1)+4\sqrt{2}G(0,1,1,2)\nonumber\\
    &\quad\,-5\sqrt{2} G(0,1,2,1))-60 G(1,0,0,0)\\
    &\quad\,-12\sqrt{30} G(1,2,1,1)-20 G(1,2,2,0))\nonumber\\
    &\quad\,+72 G(2,1,1,0)),\nonumber
\\
    \alpha_{9} &= \frac{1}{540\pi^2} \tau (72 G(0,1,1,0) (5 D_\mathrm{T} \tau  (3 G(2,1,0,1)\nonumber\\
    &\quad\,+G(0,1,0,1)+G(0,1,1,0)))\nonumber\\
    &\quad\,+\tau ^2 v_0^2( G(0,1,0,1)( (2\sqrt{2} G(0,1,1,2)\nonumber\\
    &\quad\,-7\sqrt{2} G(0,1,2,1))+170 G(0,1,1,0))\nonumber\\
    &\quad\,+30 G(0,1,0,1)^2+64 G(0,1,1,0)^2\nonumber\\
    &\quad\, -2 G(0,1,2,1)^2 +3 \sqrt{2} G(0,1,1,0) (2 G(0,1,1,2)\nonumber\\
    &\quad\,-9 G(0,1,2,1)))-2 \tau  v_0(4 \sqrt{2} G(0,1,2,1) \\
    &\quad\,\quad\, (G(1,2,2,0)-G(1,2,0,2))+G(0,1,0,1) \nonumber\\
    &\quad\,\quad\, (30 G(1,0,0,0)+3 \sqrt{30} G(1,2,1,1)\nonumber\\
    &\quad\,+14 G(1,2,2,0))+G(0,1,1,0) (90 G(1,0,0,0)\nonumber\\
    &\quad\,-20 \sqrt{3} G(1,0,1,1)+29 \sqrt{30} G(1,2,1,1)\nonumber\\
    &\quad\,+24 G(1,2,0,2)+18 G(1,2,2,0))) \nonumber\\
    &\quad\,+16 G(1,2,2,0)(2 G(1,2,0,2)-G(1,2,2,0))\nonumber\\
    &\quad\,+36 G(2,1,1,0)(G(0,1,0,1)+3 G(0,1,1,0)) ), \nonumber
\\
    \alpha_{10} &= \frac{1}{135\pi^3} \tau ^2(G(0,1,1,0)^2(\tau  v_0 ( (\sqrt{2}G(0,1,1,2)\nonumber\\
    &\quad\,+19\sqrt{2} G(0,1,2,1))-142 G(0,1,0,1))\nonumber\\
    &\quad\,+60 G(1,0,0,0)-20 \sqrt{3} G(1,0,1,1)\nonumber\\
    &\quad\,+32 G(1,2,0,2)+26 \sqrt{30} G(1,2,1,1)\nonumber\\
    &\quad\,+8 G(1,2,2,0))-34 \tau  v_0 G(0,1,1,0)^3\nonumber\\
    &\quad\,+G(0,1,1,0)(\tau  v_0 (-60 G(0,1,0,1)^2\nonumber\\
    &\quad\,+3 \sqrt{2} G(0,1,0,1) (G(0,1,1,2) \nonumber\\
    &\quad\, +5 G(0,1,2,1))+G(0,1,2,1) (4 G(0,1,2,1)\nonumber\\
    &\quad\,-3 G(0,1,1,2)))+2 (G(0,1,0,1)\\
    &\quad\,\quad\, (30 G(1,0,0,0)-10 \sqrt{3} G(1,0,1,1)\nonumber\\
    &\quad\,+7 \sqrt{30} G(1,2,1,1)+8 G(1,2,2,0))\nonumber\\
    &\quad\, + \sqrt{2}G(1,2,2,0)(G(0,1,1,2)+4 G(0,1,2,1)) \nonumber\\
    &\quad\, -8\sqrt{2} G(0,1,2,1) G(1,2,0,2)) ) \nonumber\\
    &\quad\, -G(0,1,0,1) G(0,1,1,2)\nonumber\\
    &\quad\, \quad\, (\tau  v_0 G(0,1,2,1)+2 \sqrt{2} G(1,2,2,0))), \nonumber
\\
    \alpha_{11} &= \frac{4}{135\pi^4} \tau ^3 G(0,1,1,0) ( G(0,1,1,0)^2 \nonumber\\
    &\quad\,\quad\,(36 G(0,1,0,1)-\sqrt{2} (G(0,1,1,2)\nonumber\\
    &\quad\, +4 G(0,1,2,1))) +6 G(0,1,1,0)^3\nonumber\\
    &\quad\,+G(0,1,1,0)(-3 \sqrt{2} G(0,1,0,1)(G(0,1,1,2)  \nonumber\\
    &\quad\,+2 G(0,1,2,1)) +30 G(0,1,0,1)^2 \\
    &\quad\, +G(0,1,2,1) (G(0,1,1,2)-2 G(0,1,2,1) ) )\nonumber\\
    &\quad\, +3 G(0,1,0,1) G(0,1,1,2) G(0,1,2,1)), \nonumber
\\
    \alpha_{12} &= \frac{1}{1080\pi} \tau  v_0 (-60 \tau (12 D_\mathrm{T}+\tau v_0^2) G(0,1,1,0)\nonumber\\
    &\quad\, +\tau  v_0 (\tau v_0 (12\sqrt{2} G(0,1,1,2)\nonumber\\
    &\quad\,+35\sqrt{2} G(0,1,2,1)-30 G(0,1,0,1))\nonumber\\
    &\quad\,+60 G(1,0,0,0)-40 \sqrt{3} G(1,0,1,1)\\
    &\quad\,+4\sqrt{30} G(1,2,1,1)+140 G(1,2,2,0))\nonumber\\
    &\quad\, -144 G(2,1,1,0)), \nonumber
\\
    \alpha_{13} &= \frac{1}{540\pi^2} \tau (720 D_\mathrm{T} \tau G(0,1,1,0) (G(0,1,0,1) \nonumber\\
    &\quad\,+G(0,1,1,0))+48 G(1,2,2,0)^2\nonumber\\
    &\quad\,+64 G(1,2,0,2) G(1,2,2,0)\nonumber\\
    &\quad\,+432 G(0,1,1,0) G(2,1,0,1)\nonumber\\
    &\quad\,+72 G(0,1,0,1) G(2,1,1,0)+216 G(0,1,1,0)\nonumber\\
    &\quad\,\quad\, G(2,1,1,0)+\tau^2 v_0^2(30 G(0,1,0,1)^2\nonumber\\
    &\quad\,+G(0,1,0,1)(210 G(0,1,1,0)\nonumber\\
    &\quad\,-6\sqrt{2} G(0,1,1,2)-29 \sqrt{2} G(0,1,2,1)) \nonumber\\
    &\quad\,+208 G(0,1,1,0)^2+2 G(0,1,2,1)\nonumber\\
    &\quad\,\quad\,(10 G(0,1,1,2)+3 G(0,1,2,1)) \nonumber\\
    &\quad\,-3 \sqrt{2} G(0,1,1,0) (26 G(0,1,1,2) \\
    &\quad\,+63 G(0,1,2,1)))\nonumber\\
    &\quad\, -2 \tau v_0 (G(0,1,0,1)(30 G(1,0,0,0)\nonumber\\
    &\quad\,-10 \sqrt{3} G(1,0,1,1)+\sqrt{30} G(1,2,1,1)\nonumber\\
    &\quad\,+58 G(1,2,2,0))+G(0,1,1,0)\nonumber\\
    &\quad\,\quad\, (90 G(1,0,0,0)+48 G(1,2,0,2)\nonumber\\
    &\quad\,-110 \sqrt{3} G(1,0,1,1)+23 \sqrt{30} G(1,2,1,1)\nonumber\\
    &\quad\,- 20 \sqrt{2} G(0,1,1,2)G(1,2,2,0)\nonumber\\
    &\quad\, -4 \sqrt{2}G(0,1,2,1) (2 G(1,2,0,2)\nonumber\\
    &\quad\, +3 G(1,2,2,0))+206 G(1,2,2,0)) )), \nonumber
\\
    \alpha_{14} &= -\frac{1}{135\pi^3} \tau ^2 (\tau  v_0 (60 G(0,1,1,0) G(0,1,0,1)^2 \nonumber\\
    &\quad\,+G(0,1,0,1)(194 G(0,1,1,0)^2 \nonumber\\
    &\quad\,-\sqrt{2}G(0,1,1,0) (21 G(0,1,1,2)\nonumber\\
    &\quad\,+65 G(0,1,2,1)) +7 G(0,1,1,2) G(0,1,2,1)) \nonumber\\
    &\quad\,+G(0,1,1,0) (118 G(0,1,1,0)^2\nonumber\\
    &\quad\, -\sqrt{2} G(0,1,1,0) (67 G(0,1,1,2) \nonumber\\
    &\quad\, +153 G(0,1,2,1)) +G(0,1,2,1)\nonumber\\
    &\quad\, \quad\, (41 G(0,1,1,2)+12 G(0,1,2,1))))\nonumber\\
    &\quad\, -2G(0,1,0,1)(G(0,1,1,0)(30 G(1,0,0,0)\\
    &\quad\,-30 \sqrt{3} G(1,0,1,1)+9 \sqrt{30} G(1,2,1,1)\nonumber\\
    &\quad\,+56 G(1,2,2,0)) -7 \sqrt{2} G(0,1,1,2) G(1,2,2,0))\nonumber\\
    &\quad\, - 2G(0,1,1,0)(30 G(1,0,0,0)\nonumber\\
    &\quad\, +32 G(1,2,0,2)-50 \sqrt{3} G(1,0,1,1)\nonumber\\
    &\quad\,+11 \sqrt{30} G(1,2,1,1)+68 G(1,2,2,0)) \nonumber\\
    &\quad\, +2\sqrt{2} G(0,1,1,0)(13G(0,1,1,2)\nonumber\\
    &\quad\,\quad\, G(1,2,2,0)+4 G(0,1,2,1) \nonumber\\
    &\qquad\,\, (4 G(1,2,0,2)+3 G(1,2,2,0)) ) ), \nonumber
\\
    \alpha_{15} &= \frac{4}{135\pi^4} \tau ^3 G(0,1,1,0)(30 G(0,1,1,0) G(0,1,0,1)^2 \nonumber\\
    &\quad\,+G(0,1,0,1)(-\sqrt{2} G(0,1,1,0) \nonumber\\
    &\quad\,\quad\,(11 G(0,1,1,2)+32 G(0,1,2,1))\nonumber\\
    &\quad\,+52 G(0,1,1,0)^2 +11 G(0,1,1,2)  \\
    &\quad\,\quad\,G(0,1,2,1))+G(0,1,1,0)(22 G(0,1,1,0)^2 \nonumber\\
    &\quad\, -\sqrt{2}  G(0,1,1,0)(17 G(0,1,1,2)\nonumber\\
    &\quad\,+38 G(0,1,2,1))+G(0,1,2,1) \nonumber\\
    &\quad\,\quad\,(17 G(0,1,1,2)+6 G(0,1,2,1)))), \nonumber
\\
    \alpha_{16} &= \frac{1}{135\pi^2} \tau  (\tau ^2 v_0^2 (14 G(0,1,1,0)^2+2 G(0,1,1,0) \nonumber\\
    &\quad\,\quad\, (15 G(0,1,0,1) -3\sqrt{2} G(0,1,1,2)\nonumber\\
    &\quad\,-10\sqrt{2} G(0,1,2,1)) +G(0,1,2,1) (2 G(0,1,1,2)\nonumber\\
    &\quad\,+G(0,1,2,1)))+ 2 \tau  v_0(2 \sqrt{2} G(1,2,2,0)\\
    &\quad\,\quad\,(G(0,1,1,2)+G(0,1,2,1)) \nonumber\\
    &\quad\,-G(0,1,1,0)(15 G(1,0,0,0)\nonumber\\
    &\quad\,-10 \sqrt{3} G(1,0,1,1)+4 \sqrt{30} G(1,2,1,1)\nonumber\\
    &\quad\,+20 G(1,2,2,0)))+8 G(1,2,2,0)^2), \nonumber
\\
    \alpha_{17} &= \frac{2}{135\pi^3} \tau ^2(-30 \tau  v_0 G(0,1,1,0)^3+G(0,1,1,0)^2\nonumber\\
    &\qquad\,\, (\tau  v_0 (21 \sqrt{2} G(0,1,1,2) +59 \sqrt{2} G(0,1,2,1)\nonumber\\
    &\quad\,-104 G(0,1,0,1)) +30 G(1,0,0,0)\nonumber\\
    &\quad\,-30 \sqrt{3} G(1,0,1,1) +12 \sqrt{30} G(1,2,1,1)\nonumber\\
    &\quad\,+16 G(1,2,0,2) +44 G(1,2,2,0))\nonumber\\
    &\quad\, -G(0,1,1,0)(\tau  v_0 (30 G(0,1,0,1)^2\nonumber\\
    &\quad\,-\sqrt{2} G(0,1,0,1)(3 G(0,1,1,2) \nonumber\\
    &\quad\,+17 G(0,1,2,1))+3 G(0,1,2,1) \\
    &\quad\, \quad\, (5 G(0,1,1,2)+2 G(0,1,2,1)))\nonumber\\
    &\quad\, -2 G(0,1,0,1)(-5 \sqrt{3} G(1,0,1,1)\nonumber\\
    &\quad\,+15 G(1,0,0,0)+2 \sqrt{30} G(1,2,1,1)\nonumber\\
    &\quad\, +14 G(1,2,2,0))+ 8\sqrt{2} G(0,1,2,1) \nonumber\\
    &\quad\,\quad\,G(1,2,0,2)+6\sqrt{2} G(1,2,2,0)(G(0,1,1,2)\nonumber\\
    &\quad\, +2 G(0,1,2,1)) ) -G(0,1,0,1)G(0,1,1,2) \nonumber\\
    &\quad\,\quad\, (\tau  v_0 G(0,1,2,1) +2 \sqrt{2} G(1,2,2,0))), \nonumber
\\
    \alpha_{18} &= \frac{16}{135\pi^4} \tau ^3 G(0,1,1,0) (G(0,1,1,0)^2 \nonumber\\
    &\quad\,\quad\,(19 G(0,1,0,1)-2 \sqrt{2} (2 G(0,1,1,2) \nonumber\\
    &\quad\,+5 G(0,1,2,1)))+4 G(0,1,1,0)^3 \nonumber\\
    &\quad\,+G(0,1,1,0)(15 G(0,1,0,1)^2\\
    &\quad\,-2 \sqrt{2} G(0,1,0,1)(G(0,1,1,2) \nonumber\\
    &\quad\,+4 G(0,1,2,1))+2 G(0,1,2,1)  \nonumber\\
    &\quad\,\quad\,(2 G(0,1,1,2)+G(0,1,2,1)))\nonumber\\
    &\quad\, +2 G(0,1,0,1) G(0,1,1,2) G(0,1,2,1)). \nonumber
\end{align}

\bibliography{refs}
\end{document}